\documentclass[prl,aps,twocolumn]{revtex4} 
\usepackage{psfig}
\usepackage{epsfig}
\usepackage{bm}

\newcommand{\be}{\begin{equation}}
\newcommand{\ee}{\end{equation}}

\begin{document} 
 
\title{Kinetics and scaling in ballistic annihilation} 
 
\author{Emmanuel Trizac ({\tt Emmanuel.Trizac@th.u-psud.fr)}}

\affiliation{
Laboratoire de Physique Th{\'e}orique
(UMR 8627 du CNRS), B{\^a}timent 210, Universit{\'e} de
Paris-Sud, 91405 Orsay Cedex, France 
}

\begin{abstract}
We study the simplest irreversible 
ballistically-controlled reaction, whereby particles having
an initial continuous velocity distribution annihilate upon colliding. In the 
framework of the Boltzmann equation, expressions for the exponents 
characterizing the density and typical velocity decay are explicitly worked out
in arbitrary dimension. These predictions are in excellent agreement with
the complementary results of extensive 
Monte Carlo and Molecular Dynamics simulations. 
We finally discuss the definition of universality classes indexed by a continuous
parameter for this far from equilibrium dynamics with no conservation laws.
\end{abstract} 

\maketitle

Systems with reacting particles model a rich variety of phenomena and provide
prominent situations to develop and test the foundations of non-equilibrium
statistical mechanics. In this context, the diffusion controlled first order
annihilation process ($A+A \to \emptyset$) has been extensively studied
and the corresponding decay kinetics is well understood. On the other
hand much less is known in the contrasting case where the reactants move ballistically
between the collision events, despite the relevance of such motion for 
growth and coarsening processes \cite{Krug,Gerwinski}. A few theoretical results
are available in $d=1$ dimension for such irreversible kinetic processes
with discrete initial velocity distributions. In a pioneering work,
Elskens and Frisch show from combinatorial considerations that the particle 
density $n(t)$ decays like $1/\sqrt{t}$ for the simplest binary velocity 
distribution \cite{Elskens}.
Powerful generalizations of this result were obtained still in 1D, either 
for a larger class of stochastic ballistic annihilation and coalescence models
\cite{Blythe,Kafri} or from kinetic theory for discrete multi-velocity
distributions \cite{Piasecki,Droz}. No exact results could be obtained for
the generic case of
continuous distributions, where the decay exponents have been computed 
numerically \cite{BenNaim,Rey,Krapivsky}. Recently however, 
Krapivsky and Sire considered the latter situation in the framework of the 
Boltzmann equation (relying on the so-called ``molecular 
chaos'' factorization \cite{Resibois})
and derived bounds for the exponents as well as their leading
large $d$ behavior. The existing body of literature has essentially
focussed numerically on the one dimensional case, and no accurate 
predictions seem to be available for the decay exponents.

In this Letter, we obtain predictions for the decay exponents and velocity
distribution (assumed initially continuous), 
revisiting Boltzmann kinetic theory in arbitrary dimension, 
with the explicit inclusion of non Gaussian corrections to velocity 
distributions. These predictions are
compared both with the existing numerical results in 1D and the 
expressions derived in \cite{BenNaim,Krapivsky}, and further tested against
extensive numerical simulations in dimension 2 and 3, following two 
complementary routes: we first solve the mean-field non-linear Boltzmann 
equation describing the annihilation process by means of a Monte Carlo
scheme, which validates the analytical expressions obtained within the 
molecular chaos framework; second, we go beyond mean-field and investigate the
exact decay kinetics by implementing Molecular Dynamics simulations. 
The two numerical approaches yield the same exponents, in excellent agreement
with the analytical prediction. Finally, we address the question of universality
in this process \cite{Rey}
by partitioning the 
possible continuous velocity distributions into
groups associated with the same asymptotic dynamic scaling behaviour, akin
to equilibrium universality classes. 

We consider an assembly of identical spherical particles with 
radius $\sigma$ in dimension 
$d$, with initial velocity distribution $f({\bm v},t=0)$ and random initial 
positions. 
Particles
follow free flight motion until a collision occurs which results in the removal of
both partners. We are interested in the time evolution of density 
$n(t) = \int f({\bm v},t) d{\bm v}$ and typical velocity $\overline v(t)$, 
related to the 
kinetic temperature $T(t)$ defined as the variance of the velocity distribution
\be
T(t) = \frac{1}{n(t)}\,\int v^2 \,f({\bm v},t) \, d{\bm v} \,= \,(\overline v)^2.
\ee
Insight into the decay kinetics may be gained by writing the rate equations for 
$n$ and $T$
\begin{eqnarray}
&&\frac{d n}{dt} = - \omega(t)\, n
\label{eq:alphadef1}\\
&&\frac{d (nT)}{dt} = -\omega(t) n T_{\hbox{\scriptsize coll}} 
\label{eq:alphadef2}
= - \alpha\, \omega(t)\, n\, T,
\end{eqnarray}
where the first line stands for a definition of the instantaneous 
mean collision frequency $\omega$, while $T_{\hbox{\scriptsize coll}}$
is the time dependent total kinetic energy of a colliding pair, 
which is thus dissipated in a binary encounter, as stated by the rhs equality 
in Eq. (\ref{eq:alphadef2}). 
On dimensional grounds, the collision frequency is expected to scale like the 
inverse time, which together with Eqs. (\ref{eq:alphadef1}) and (\ref{eq:alphadef2})
implies an algebraic time decay for $n$ and $\overline v$, as well as a 
time-independent energy dissipation parameter $\alpha$ [defined in Eq.
(\ref{eq:alphadef2}) as $\alpha=T_{\hbox{\scriptsize coll}}/T$]. 
We therefore introduce two exponents 
$\xi$ and $\gamma$ such that $n(t) \propto t^{-\xi}$
and $\overline v \propto t^{-\gamma}$ (and $T \propto t^{-2 \gamma}$).  With a 
ballistic dynamics controlled by the mean-free-path $\ell \propto 1/(n \sigma^{d-1}$),
the collision frequency may be written as the ratio $\overline v / \ell$. 
From $\omega \propto 1/t$, we obtain the scaling relation $\xi+\gamma=1$ 
\cite{BenNaim,Rey,Krapivsky,Dufty}, which may be combined with 
the ratio of Eqs. (\ref{eq:alphadef1}) and (\ref{eq:alphadef2}) 
to give
\be
\xi = \frac{2}{1+\alpha} \quad \hbox{ and } \quad 
\gamma = \frac{\alpha-1}{\alpha+1}.
\label{eq:xialpha}
\ee
Since particles with a higher velocity are likely to disappear with 
a higher rate than the average particle with temperature $T$, we expect 
$\alpha=T_{\hbox{\scriptsize coll}}/T$ to be larger than 1,
so that the typical velocity should decrease with time [$\gamma >0$
from Eq. (\ref{eq:xialpha})]. 
This moreover explains the failure of the naive 
mean-field picture where the density decay rate is written 
$\dot n \propto - n^2$, so that $n(t) \propto 1/t$. This 
transparency limit would hold in the absence of collisional correlations
($\alpha=1$) which becomes only asymptotically exact in the limit of infinite
dimension $d$. 

We now turn to the computation of $\alpha$ within the molecular 
chaos framework, which is {\it a priori} an uncontrolled approximation.
It will however be shown to capture the essential
collisional correlations missed by the naive mean-field argument,
and to provide decay exponents in excellent agreement with their
numerical counterparts. 
The corresponding Boltzmann equation reads
\be
\frac{\partial f({\bm v},t)}{\partial t}\,=\,
-f({\bm v},t)\,\int d{\bm w} |{\bm v}-{\bm w}|\, f({\bm w},t),
\label{eq:Boltzmann}
\ee
which implies that if the initial distribution behaves like a power law
$|v|^\mu$ near the velocity origin, this property is preserved at subsequent 
times by the dynamics, which in turn should affect the exponents $\xi$ 
and $\gamma$, expected to depend explicitly on $\mu$ (as appears on the
analytical predictions of Ben-Naim et al \cite{BenNaim} 
$\xi = (2d + 2\mu)/(2d + 2\mu+1)$, or on the bounds derived by Krapivsky and Sire
\cite{Krapivsky}). Looking for a scaling solution of the kinetic equation
(\ref{eq:Boltzmann}), we introduce a rescaled velocity 
${\bm c} = {\bm v}/\overline v$
and rescaled single particle distribution function $\varphi$ through
\be
f( {\bm v},t) \, = \,\frac{n(t)}{\overline v^{d}}\,\, \varphi({\bm c},t),
\label{eq:scaling}
\ee
so that $\varphi({\bm c},t)$ is the probability distribution function of 
the velocity ${\bm c}$ at time $t$, satisfying the constraints 
$\int \varphi \,d{\bm c} = 1$ and $\int c^2 \varphi \,d{\bm c}= 1$
at any time. If $f( {\bm v},t) $ evolves into a self-similar decay 
state, the only relevant time dependence occurs via $n(t)$ and $\overline v(t)$,
so that $\varphi({\bm c},t)$ no longer depends on time and the evolution equation for 
$f$ (assumed isotropic) translates into  
\begin{eqnarray}
\left[1 + 
\left(\frac{1-\alpha}{2}\right)\left(d+c_1\frac{d}{dc_1}\right)\right]\,
\varphi( c_1) \,=\,&&\nonumber\\
\varphi(c_1) \int\! d{\bm c_2}\, 
\frac{c_{12}}{\langle c_{12}\rangle}\, \varphi (c_2), &&
\label{eq:Boltzrescaled}
\end{eqnarray}
where $\langle (\ldots) \rangle = \int (\ldots) 
\varphi(c_1) \varphi(c_2)d{\bm c}_1 d{\bm c}_2$
so that 
$\langle c_{12} \rangle \,\equiv \langle|{\bm c}_1\!-\!{\bm c}_2|\rangle$
denotes the rescaled collision frequency.

Equation (\ref{eq:Boltzrescaled})
may be considered as an eigenvalue problem for $\alpha$, which has been 
computed numerically in 1D \cite{Krapivsky}. However, it is useful to reformulate
Eq. (\ref{eq:Boltzrescaled}) into an infinite hierarchy of consistency relations
obtained by computing the corresponding moment of order $p$:
\be
\alpha \,=\, 1 \,+\,\frac{2}{p}\left(
\frac{\langle c_{12}\,c_1^p\rangle}{\langle c_{12}\rangle\langle c_1^p\rangle}-1
\right).
\label{eq:hierarchy}
\ee
Note that the special case $p=2$ 
coincides with the definition of $\alpha$ through the kinetic energy dissipation
as expressed by Eq. (\ref{eq:alphadef2}): 
$\alpha \,=\, T_{\hbox{\scriptsize coll}}/T = 
\langle c_{12}\,c_1^2\rangle/(\langle c_{12}\rangle\langle c_1^2\rangle)$.
We look for explicit solutions by expanding $\varphi$ in a basis of Sonine
functions \cite{Landau}
\be
\varphi({\bm c}) \, = \, {\cal M}({\bm c})\, \left[
1 + \sum_{n=1}^\infty a_n\, S_n(c^2)
\right]
\label{eq:sonine}
\ee
where the polynomials $S_n$ are orthogonal with respect to the
Gaussian weight ${\cal M}({\bm c})$. Computing the averages
involved in (\ref{eq:hierarchy}) from the functional expression 
(\ref{eq:sonine}) provides a system of equations for the coefficients
$a_n$.

In practice, only a few terms are required in the expansion 
(\ref{eq:sonine}) in order to get a precise estimation 
for $\alpha$, provided relations of lowest
order $p$ as possible are retained among the hierarchy (\ref{eq:hierarchy}). 
In this respect, taking the limit of vanishing velocity of (\ref{eq:Boltzrescaled})
yields the ``optimal'' relation involving $\alpha$ and moments of $\varphi$ of order
1:
\be 
\alpha = 1+ \frac{2}{\mu + d}\left(
1-\frac{\langle c_1 \rangle}{\langle c_{12}\rangle}\right),
\label{eq:alphautile}
\ee
that we consider as the first equation of (\ref{eq:hierarchy}) corresponding to
the limit $p \to 0^+$. At Gaussian order for $\varphi$ [i.e. truncating
(\ref{eq:sonine}) at order $n=0$], it is straightforward to get
\be
\alpha \,=\, \alpha_0 \,=\, 1 + \frac{2}{d+\mu}\left(1-\frac{\sqrt{2}}{2}\right) 
\label{eq:alpha0}
\ee
which, together with Eq. (\ref{eq:xialpha}) yields the zeroth order
estimation for $\xi$:
\be
\xi_0 \,=\, \frac{2 d + 2\mu}{2(d+\mu+1) -\sqrt{2}}.
\label{eq:xi0}
\ee
It is noteworthy that in the limit of large dimension, we obtain
$\xi_0 \sim 1-d^{-1}(1-1/\sqrt{2}) + {\cal O}(1/d^2)$ irrespective of $\mu$,
which has been shown to be the exact $1/d$ behaviour within Boltzmann
molecular chaos framework \cite{Krapivsky}. 
The first non Gaussian correction is carried by $a_2$ ($a_1$
identically vanishes from the definition of temperature \cite{vanNoije}) 
and this coefficient is related to the kurtosis of the velocity distribution: 
$a_2$ is proportional to the fourth cumulant 
$\langle c_i^4 \rangle - 3 \langle c_i^2 \rangle^2 $, where $c_i$ is a given 
Cartesian coordinate of ${\bm c}$. After a lengthy calculation performed
at linear order in $a_2$, we obtain 
\begin{eqnarray}
&& a_2 \, =  \,8 \,\frac{ \mu + d (3 -2 \sqrt{2})}
{ 4 d^2 + 6 \mu + d (6 + 4 \mu -\sqrt{2})} 
\label{eq:a2} \\
&& \alpha_2 \, = \, \alpha_0 \, + \, \frac{\sqrt{2}}{16 d} \, a_2. 
\label{eq:alpha2}
\end{eqnarray}

The above predictions rely on a perturbative expansion starting from the
Maxwellian ${\cal M}$ (regular at ${\bm v}={\bm 0}$) and are therefore
expected to be particularly relevant for $\mu$ close to 0. The agreement with the
existing numerical data is excellent; an accurate estimation has been reported 
in 1D within molecular chaos for the much studied $\mu=0$ case \cite{Krapivsky}: 
$\xi = 0.769(5)$ whereas we obtain 
at zeroth order $\xi_0 = 0.773$ from (\ref{eq:xi0}) and at second order 
$\xi_2 = 2/(1+\alpha_2) = 0.769(3)$ from Eq. (\ref{eq:alpha2}). This exponent
is compatible with its counterpart extracted from the exact dynamics (0.78 in
\cite{Rey}). Moreover, we have investigated numerically the annihilation dynamics 
in higher dimensions by means of {\em a)} the Direct Simulation Monte Carlo 
procedure \cite{Bird} (DSMC) solving the non-linear homogeneous Boltzmann equation 
(\ref{eq:Boltzmann}) and {\em b)} Molecular Dynamics simulations (MD) implementing
the exact dynamics with periodic boundary conditions \cite{Allen}. The DSMC
technique provides precise data for the velocity distributions and decay exponents,
and allows to test the validity of the analytical truncated expansion of the 
scaling form $\varphi$, leading to (\ref{eq:alpha0}) or (\ref{eq:alpha2}).
Alternatively, MD results assess the reliability of the molecular chaos ansatz,
but are more demanding on computer resources: on the one hand, the system 
needs to reach very low densities in order to develop the self-similar decay stage
where $f({\bm v},t)$ takes the scaling form (\ref{eq:scaling}), but on the other 
hand, the mean free path $\ell$ which increases with time like $t^\xi$ 
must remain smaller than the simulation box size $L$, which provides a lower bound
for $n(t)$ or equivalently an upper bound for accessible times before
finite size effects hinder the precise determination of $\xi$ and $\gamma$. 
In practice, we considered systems with $N=10^5- 5\,10^5$ particles 
in MD and $N=10^6-10^7$ in DSMC where it is further possible to average over $10^3$ to 
$10^4$ replicas to increase the statistics of the
velocity distributions, which is crucial for computations at large times 
with a concomitant low number of particles left.

\begin{center}
\begin{figure}[thb]
\epsfig{figure=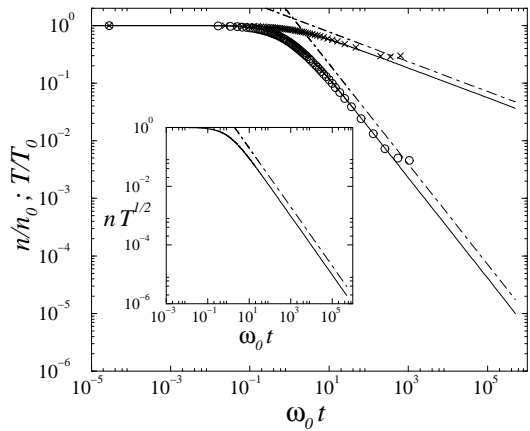,width=6.9cm,angle=0}
\vskip -3mm
\caption{Evolution of the density (lower sets) and kinetic temperature (upper sets),
normalized by their initial values. At $t=0$, the velocity distribution 
is Maxwellian ($\mu=0$), with a collision frequency denoted $\omega_0$. 
MD results are shown by symbols (circles for $n$ and crosses
for $T$) and DSMC by continuous curves. The dashed lines have slopes given by 
the theoretical predictions. Inset: check of the scaling relation 
$\xi+\gamma=1$ where $n\sqrt{T}$ is expected to scale like $t^{-\xi-\gamma}$;
the dashed line has slope -1.
\label{fig:fig1}}
\end{figure}
\end{center}

\vskip -6mm
\begin{table}[hbt]
\begin{center}
\begin{tabular}{|c|ccc|}
\hline
({\bf 1D}) values of $\mu$~~  & ~~~~$-4/5$~~~~    & ~~~~$-1/2$~~~~ & ~~~~$0$~~~~\\
\hline
Prediction \cite{BenNaim,Krapivsky} & 0.28 & 0.5  & 0.666   \\
Simulation  \cite{BenNaim}      & 0.32/0.37 & 0.56/0.60 &  0.769   \\
$\xi_2$ from Eq. (\ref{eq:alpha2}) & 0.32  & 0.60  &  0.769   \\
\hline
\end{tabular}
\caption{Decay exponent $\xi$ in 1 dimension.
\label{tab:tab1}}
\end{center}
\end{table}
\vskip -5mm
\begin{table}[hbt]
\begin{center}
\vskip -5mm
\begin{tabular}{|c|cccc|}
\hline
({\bf 2D}) values of $\mu$~~  
     &~~$-1$~~~&~~~$-1/2$~~~&~~~~$0$~~~~&~~~$3$~~~\\
\hline
Prediction  \cite{BenNaim,Krapivsky} 
        & ~0.66 & ~~0.75 & 0.800 & ~~~0.91~ \\
Simulation   
        & ~0.75 & ~~0.83 & 0.870 & ~~~0.97~ \\
$\xi_2$ from Eq. (\ref{eq:alpha2})
        & ~0.76 & ~~0.84 & 0.870 & ~~~0.95~ \\
\hline
\end{tabular}
\caption{Exponent $\xi$ in 2D; the simulation data are the Monte Carlo results of
the present work.
\label{tab:tab2}}
\end{center}
\end{table}

\vskip -5mm
The results of two dimensional
simulations are shown in Fig. \ref{fig:fig1} where it appears that the MD data 
are fully compatible with DSMC, although less precise. 
For $\omega_0 t \simeq 10^3$, the MD density and temperature tend to saturate,
which corresponds to the upper time limit where $\ell \simeq L$, and the subsequent
evolution is discarded. The predictions $\xi_0 = 0.872$ and $\xi_2=0.870$ 
for $\mu=0$
(indistinguishable in Fig \ref{fig:fig1}) are in good agreement with 
the simulations, irrespective of the initial $f({\bm v})$ chosen 
(we considered several distributions with the constraint 
$\mu=0$, see below the discussion concerning universality). 
The above exponent is compatible with that reported in the 
context of a multi-particle lattice gas method (0.87 \cite{Chopard}). Moreover, 
the initial spatial configuration is irrelevant (the long time dynamics
and rescaled velocity distributions
are the same starting from a fluid-like structure or from various
crystalline arrays), and the scaling relation $\xi+\gamma=1$ is seen to be well 
obeyed in the asymptotic regime (inset of Fig. \ref{fig:fig1}). The same 
scenario holds in dimensions 3 and 4, where the predictions at zeroth and second
order are very close, and indistinguishable from the numerics 
($\xi_0 \simeq 0.91$ in 3D and 0.93 in 4D for $\mu=0$). 
However, the agreement is expected
to become worse as $\mu$ deviates from 0 (with $\mu>-d$ to 
ensure proper normalization). This is confirmed in
Tables \ref{tab:tab1} and \ref{tab:tab2} which summarize the 
results obtained for various
$\mu$, with comparison to the theoretical prediction
of Ben-Naim et al. \cite{BenNaim} (coinciding with the lower bound
for $\xi$ obtained in \cite{Krapivsky}, the upper bound being 1). 
For $\mu=0$, the non Gaussian parameter $a_2$ is small [with an
even smaller correction to $\alpha$ due to the prefactor $\sqrt{2}/(16 d)$ 
in (\ref{eq:alpha2})]. This fourth cumulant however rapidly increases with $\mu$,  
so that inclusion of higher order terms [$n=3$\ldots in (\ref{eq:sonine})] 
would be required
to obtain the same level of accuracy as for regular distributions near the 
velocity origin.

\vskip -4mm
\begin{center}
\begin{figure}[htb]
\vspace{-0.5cm}
\epsfig{figure=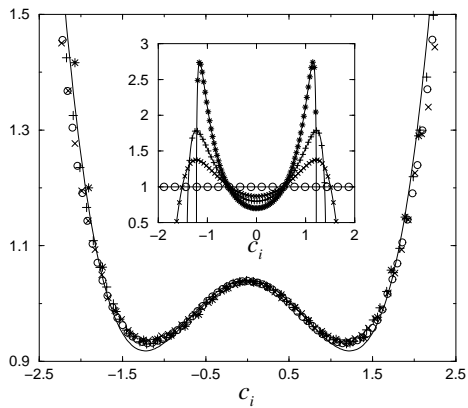,width=6.5cm,angle=0}
\vskip -5mm
\caption{Plots of $\varphi({\bm c}_i)/{\cal M}({\bm c}_i)$ versus $c_i$
in 2D. The inset shows 4 different initial distributions with $\mu=0$, 
one of them being Gaussian [thus corresponding to the
flat curve (circles)]. These distributions having very different $a_2$ at $t=0$ 
collapse onto a master curve in the asymptotic scaling regime (main graph). 
The thick curve
is the prediction $1+a_2 S_2(c_i^2)$ where $a_2$ is given by Eq. (\ref{eq:a2})
and $S_2(x) = x^2/2-3x/2+3/8 $.
The symbols (stars, crosses, pluses and circles) refer to the same distributions 
at late times (main graph) and at $t=0$ (inset). The results have been obtained
by averaging over $10^4$ replicas of a system with $N=5\, 10^6$ particles.
\label{fig:fig2}}
\end{figure}
\end{center}

\vskip -8mm
In the remainder, we consider the possibility to define universality classes
for ballistic annihilation kinetics, in the following sense: does $\mu$ completely 
specify the asymptotic velocity distribution and decay exponents,
irrespective of further details concerning the initial conditions \cite{Rey}?
To answer this question we have run several simulations (MD and Monte Carlo) 
corresponding to different initial conditions sharing the same $\mu$, 
for several values of this parameter. The corresponding decay exponents
$\xi$ and $\gamma$ are monitored, which provides a first test, however quite 
insensitive to possible non Gaussianities (see above the numerical proximity
between $\xi_0$ and the non Gaussian corrected $\xi_2$). A more
sensitive and severe probe is provided by the kurtosis $a_2$, 
which may be computed in two
different ways: first from its definition involving the fourth cumulant
$\langle c_i^4 \rangle - 3 \langle c_i^2 \rangle^2 $, or alternatively 
from the direct computation of $\varphi({\bm c})/{\cal M}({\bm c})$, 
which may further be compared to the analytical expansion $1 + a_2 S_2(c^2)$
with $a_2$ given by Eq. (\ref{eq:a2}) (recall that $a_1 \equiv 0$). 
The latter method is illustrated in Fig. \ref{fig:fig2} where the four initial
distributions shown in the inset evolve after a transient towards the same
attractor, that is furthermore in quantitative agreement with the Sonine prediction. 
Moreover the same values of $\xi$ and $\gamma$ are measured within 
statistical inaccuracy for the 4 distributions.
We have observed the same phenomenology for $\mu \neq 0$, which points to the 
relevance of defining universality classes of initial conditions 
as distributions having the same regularity exponent $\mu$, as conjectured in 1D
for $\mu=0$ \cite{Rey}.

In conclusion, we have shown that the non trivial dynamic scaling behaviour 
of ballistic annihilation may be investigated within Boltzmann 
kinetic theory, and accurate decay exponents have been explicitly worked out. 
Their evaluation (\ref{eq:xi0}) at zeroth order turns out to be straightforward, 
but follows from a kinetic equation and is therefore specific to the precise
model considered here. A more versatile approach that would apply to any 
ballistically controlled reaction (including coalescence with arbitrary
conservation laws, with or without stochasticity in the reactions) consists
in reconsidering the rate equations (\ref{eq:alphadef1}) and (\ref{eq:alphadef2}),
and identify the proper energy dissipation parameter $\alpha$ before approximating it
assuming a Gaussian velocity distribution.  This 
``model-independent'' approach gives
$\alpha=1+1/(2d)$ in the particular case of pure annihilation, 
which corresponds to $\xi = 4d/(4d+1)$
(i.e. 0.8, 0.89 and 0.92 in dimensions 1, 2 and 3)
in reasonable agreement with the exponents mentioned above
(0.77, 0.87 and 0.91 respectively). We conjecture that 
the exponent $\xi= 4d/(4d+1)$
becomes exact when the particles annihilate with probability $p$ 
(and collide elastically otherwise), 
in the limiting case $p \to 0^+$
(whereas $p=1$ for ``pure'' annihilation). 
This hopefully provides an
illustration of the central role played by the energy dissipation parameter
$\alpha$ in ballistically controlled reactions, and calls for further 
investigations
with more involved reactions.
\null


\vskip -7mm 


\begin{thebibliography}{99}
\null\vskip -3mm
\bibitem{Krug}
\null\vspace{-6mm}
  J. Krug and H. Spohn, 
  Phys. Rev. A {\bf 38}, 4271 (1988). 
  
\bibitem{Gerwinski}  
  M. Gerwinski and J. Krug, 
  Phys. Rev. E {\bf 60}, 188 (1999).

\bibitem{Elskens}
  Y. Elskens and H.L. Frisch,
  Phys. Rev. A {\bf 31}, 3812 (1985).

\bibitem{Blythe}  
  R.A. Blythe, M.R. Evans and Y. Kafri,
  Phys. Rev. Lett. {\bf 85}, 3750 (2000).

\bibitem{Kafri}
  Y. Kafri,
  J. Phys. A: Math. Gen. {\bf 33}, 2365 (1999).   

\bibitem{Piasecki}
  J. Piasecki,
  Phys. Rev. E {\bf 51}, 5535 (1995).
  
\bibitem{Droz}
  M. Droz, P.A. Rey, L. Frachebourg and J. Piasecki,
  Phys. Rev. Lett. {\bf 75}, 160 (1995).

\bibitem{BenNaim}  
  E. Ben-Naim, S. Redner and F. Leyvraz, 
  Phys. Rev. Lett. {\bf 70}, 1890 (1993); 
  E. Ben-Naim, P.L.  Krapivsky, F. Leyvraz and S. Redner,
  J. Phys. Chem. {\bf 98}, 7284 (1994). 

\bibitem{Rey}
  P.A. Rey, M. Droz and J. Piasecki, 
  Phys. Rev. E {\bf 57}, 138 (1998).  
  
\bibitem{Krapivsky}
  P.L. Krapivsky and C. Sire, 
  Phys. Rev. Lett. {\bf 86}, 2494 (2001).

\bibitem{Resibois}
  P. R\'esibois and M. de Leener, 
  {\it Classical Kinetic Theory of Fluids}, John Wiley and Sons (1977).

\bibitem{Dufty} 
  Incidentally, the relation $\xi+\gamma =1$ equally
  applies to the homogeneous cooling stage of (inelastic) granular matter 
  where density in conserved ($\xi=0$) so that $T \propto t^{-2}$,
  as observed, see e.g. 
  J. Dufty,
  cond-mat/0108444 and references therein.
   
\bibitem{Landau}
  L. Landau and E. Lifshitz, 
  {\it Physical Kinetics}, Pergamon Press (1981).
  
\bibitem{vanNoije}
T.P.C. van Noije and M.H. Ernst,  Gran. Matter {\bf 1}, 57 (1998).
   
\bibitem{Bird}
  G. Bird, ``Molecular Gas Dynamics'' (Oxford University Press, New York, 1976)
  and ``Molecular Gas Dynamics and the Direct Simulation of Gas flows''
 (Clarendon Press, Oxford, 1994).

\bibitem{Allen}
  M.P. Allen and D.J. Tildesley
  ``Computer Simulations of Liquids'' (Clarendon Press, Oxford, 1987).

\bibitem{Chopard}
  B. Chopard, A. Masselot and M. Droz,
  Phys. Rev. Lett. {\bf 81}, 1845 (1998). 

\end{thebibliography}
\end{document}